# Atomic ironing for conformal interfaces in freestanding oxide nanomembranes

Jeongkeun Song[1], Keaghan Knight[2], Keun-Yeol Park[3], Young-Hoon Kim[4], Tyler J. Sachs[1,5], Hu Miao[1], Jiaqiang Yan[1], Celesta S. Chang[3,6], Ambrose Seo[2], and Ho Nyung Lee[1*]

*[1] Materials Science and Technology Division, Osak Ridge National Laboratory, Oak Ridge, TN 37831, USA.*

*[2] Department of Physics and Astronomy, University of Kentucky, Lexington, KY 40506, USA.*

*[3]Department of Physics and Astronomy, Seoul National University, Seoul, 08826, Republic of Korea.*

*[4] Center for Nanophase Materials Sciences (CNMS), Oak Ridge National Laboratory (ORNL), Oak Ridge, TN 37831, USA.*

*[5]Department of Physics, The University of Texas at Austin, Austin, Texas 78712, USA.*

*[6]Institute of Applied Physics, Seoul National University, Seoul, 08826, Republic of Korea.*

Correspondence to: hnlee@ornl.gov

**Abstract**

Atomically well-defined interfaces are essential for realizing functional interlayer coupling in stacked and twisted oxide heterostructures, yet transferred freestanding oxides can develop structural inhomogeneities, including local curvatures, interfacial voids, and incomplete contacts with substrates. Here, we demonstrate that these transfer-induced structural disorders and imperfections evolve substantially and can be suppressed by post-transfer thermal treatments, yielding bulk-like crystalline freestanding oxide nanomembranes. Using freestanding $SrTiO_3$ membranes as a model system, we report a systematic understanding on the role of “atomic ironing” and the associated evolution of the lattice mosaicity, interfacial roughness, and optical bandgap. We further demonstrate atomically bonded conformal interfaces in twisted $SrTiO_3$ heterostructures, highlighting the importance of post-transfer thermal processes. These results establish a synthetic strategy to restore the structural integrity and intrinsic bulk-like functionality in freestanding oxide nanomembranes, enabling stacked, twisted, and heterogeneously integrated oxide architectures with atomically well-defined interfaces for emerging quantum functionalities.

## Introduction

The emergence of cleavable or weakly bonded two-dimensional (2D) materials and freestanding membranes has transformed the design of electronic and quantum systems by enabling heterostructures beyond conventional thermodynamic and epitaxial constraints. In particular, stacking, twisting, and stretching atomically thin layers introduce new structural degrees of freedom that can generate moiré superlattices, correlated electronic states, and tunable optical responses[1-7] that are absent in their bulk counterparts. Extending these concepts to complex oxides is particularly attractive because their intertwined charge, spin, orbital, and lattice degrees of freedom offer a broader landscape of emergent functionalities[8-15]. However, unlike van der Waals materials, most oxides cannot be readily exfoliated because of their strong ionic and covalent bondings[16,17], limiting access to stacked artificial oxide heterostructures.

Recent advances in remote epitaxy and sacrificial-layer liftoff techniques have enabled the fabrication of freestanding oxide membranes, providing a pathway to overcome this limitation[18-23]. By releasing epitaxial oxides from their parent substrates, these membranes can be heterogeneously integrated with other materials platforms, including silicon and flexible supports, enabling deliberate manipulation through bending[24,25], stacking[26], and twisting[27-31]. Moreover, such approaches further enable the monolithic three-dimensional integration of diverse oxide functionalities, including ferroelectricity, magnetism, ionicity, and superconductivity, into microelectronic and quantum-device architectures. Thus, freestanding oxide membranes have emerged as a promising platform for a wide range of technological applications, including flexible electronics, oxide photonics, and moiré heterostructures with novel quantum functionalities.

Realizing this potential, however, requires precise control of the membrane interface after release and transfer. In contrast to the dry-transfer processes commonly used for van der Waals materials[32], sacrificial-layer-based release generally involves wet chemical etching, during which freestanding oxides can develop microscopic and macroscopic imperfections, such as wrinkles, local curvature, interfacial voids, and incomplete contact with the underlying substrate[21,33,34]. Their effects are particularly consequential in stacked and twisted oxide heterostructures, where atomically flat and conformal interfaces are essential for controlling interlayer coupling, local atomic registry, charge transfer, and moiré potentials. Yet how transferred oxide membranes can be reconstructed toward atomically conformal interfaces remains poorly understood.

Here, using freestanding $SrTiO_3$ membranes (m-STO) as a model system, we show how the structure and property of oxide membranes, including membrane curvature, lattice mosaicity, interfacial roughness, and optical bandgap, evolve upon a post-transfer thermal treatment, yielding atomically flat, conformal membrane-substrate interfaces. This structural improvement restores the intrinsic optical response without introducing detectable low-energy in-gap states. We further demonstrate that a proper thermal treatment eliminates discernible interfacial gaps in a twisted m-STO heterostructure.

**Results and Discussion**

To examine the structural changes introduced during membrane release and transfer, we first fabricated freestanding m-STO using a sacrificial-layer-based approach. STO/$Sr_4Al_2O_7$ (SAO) thin films were epitaxially grown on (001)-oriented $TiO_2$-terminated STO substrates by pulsed laser deposition. The SAO sacrificial layer was then selectively dissolved in water, releasing the epitaxial m-STO from the parent substrate (Figure 1a–c). The released m-STO was successfully transferred onto a (001)-oriented $SiO_2$/Si substrate (Figure 1d). The transferred membrane exhibits a continuous and well-defined square geometry without macroscopic tearing or delamination over the $SiO_2$/Si substrate, indicating the successful release and transfer of the m-STO. During release and transfer, however, freestanding membranes can develop structural inhomogeneities[21,33], such as local curvature, wrinkles, and interfacial voids, resulting in non-uniform membrane–substrate contacts (Figure 1b). To mitigate these transfer-induced disorders, thermal annealing was employed to promote atomic interface formation between the transferred membrane and the receiving substrate (Figure 1c).

The structural properties of the membranes were first examined by X-ray diffraction (XRD). As shown in Figure 1e, the as-grown STO/SAO/STO heterostructure exhibits sharp diffraction peaks with pronounced Laue oscillations, indicating high-crystalline quality. After transfer, m-STO on both $SiO_2$/Si and $Al_2O_3$ retain the characteristic STO 001 and 002 reflections near 22.8° and 46.5°, respectively, confirming preservation of the structural integrity and crystallinity. However, the diffraction features become significantly broadened compared to the as-grown thin film, suggesting increased structural disorder. Consequently, XRD rocking curve measurements (Figure 1f) reveal a substantial increase in full width at half maximum (FWHM) from ~ 0.08° in the as-grown state to ~ 1.0° after transfer. This peak broadening indicates enhanced mosaicity arising from local bending, lattice tilt variations, and wrinkle formation introduced during the release and transfer processes. In other words, these results indicate that although the SAO-assisted process enables successful release and transfer of crystalline m-STO, the transfer process introduces pronounced structural inhomogeneities[19,35] that require additional processes to recover atomically uniform membrane–substrate interface.

We next examine the effect of thermal annealing on structural recovery. Figure 2a displays XRD 2θ– θ scans of m-STO annealed at different temperatures ranging from 300 to 700 °C. While the position of the 002 reflection remains nearly unchanged, the diffraction features become progressively pronounced with increasing annealing temperature, suggesting improved crystallographic coherence and reduced structural disorder. Low-angle X-ray reflectivity (XRR) measurements further reveal the systematic evolution of membrane thickness and surface/interface roughness. As shown in Figure 2b, the XRR oscillations become more pronounced and persist to higher angles with increasing annealing temperature, consistent with improved thickness uniformity and reduced roughness. Quantitative fitting using a multilayer model shows that the effective membrane thickness decreases from ~13.9 nm at 300 °C to ~13.0 nm at 600 °C, while the interface roughness is substantially reduced from ~4.4 to ~2.9 Å.

Rocking curve analysis of the STO 002 reflection further confirms the annealing-induced structural recovery (Figure 2c). The rocking-curve FWHM decreases from ~1.0° in the as-transferred state to ~0.1° after annealing, approaching the crystallinity of the as-grown film. Figure 2d summarizes the evolution of effective membrane thickness, interface roughness, and rocking-curve FWHM as a function of annealing temperature. Collectively, these results demonstrate that post-transfer thermal annealing effectively recovers the structural uniformity of m-STO by reducing interfacial roughness, mosaicity, and lattice misorientation while improving the membrane–substrate contact.

We next examined whether this structural recovery is accompanied by a corresponding recovery of the intrinsic optical response of the m-STO. STO is a wide bandgap oxide (~ 3.2 eV) exhibiting a strong optical absorption in the ultraviolet region associated with O–2*p* to Ti–3*d* electronic transitions. Figure 3a illustrates the transmission measurement geometry for m-STO samples on double-side polished $Al_2O_3$ substrates. Transmission spectrum (Figure 3b) show that while the visible-range transmission remains nearly unchanged, the ultraviolet transmission decreases systematically with increasing annealing temperature. The absorption edge becomes sharper and approaches that of single-crystalline STO, indicating an improved optical response.

These observations indicate a correlation between improved structural uniformity and the recovery of the intrinsic optical response of m-STO. Note that we have measured transmission spectrum of the polymer support on double-side polished $Al_2O_3$ substrates to compare with that of STO, which shows distinctive transmission spectra allowing it to be selectively excluded from the data (Figure S1). The extracted absorption coefficient spectra shown in Figure 3c reveal distinct features near 3.8, 4.2, and 4.8 eV, corresponding to intrinsic interband transitions[36-38]. Notably, the absorption coefficient increases with increasing annealing temperature, particularly in the 4 – 5 eV range, while the peak positions remain unchanged. This behavior indicates that thermal annealing enhances the strength and clarity of intrinsic optical transitions without altering the underlying electronic structure.

To confirm the enhancement of optical absorption without changes in electronic structure, Tauc analysis was performed (Figure 3d and Figure S2). Despite the substantial increase in absorption intensity, the optical band gap remains nearly constant at ~3.2 eV. This indicates that thermal annealing enhances the optical transition strength without substantially modifying the intrinsic band structure of STO. Consistently, ellipsometry measurements show that the dielectric response of annealed membranes approaches that of bulk STO (Figure S3). Figure 3e summarizes the evolution of the absorption coefficient and band gap with annealing temperature. While the absorption coefficient increases monotonically, the optical band gap remains essentially unchanged, showing that structural recovery is accompanied by enhanced optical transitions without substantial modification of the underlying electronic structure.

To further examine the annealing-induced structural recovery and the membrane-substrate interface at the atomic scale, we performed cross-sectional high-angle annular dark-field scanning transmission electron microscopy (HAADF-STEM) on as-transferred and annealed m-STO on $SiO_2$/Si, viewed along the STO [110] zone axis. The as-transferred m-STO exhibits pronounced structural non-uniformity, including local bending and spatially varying contrast, indicating that the local orientation deviates from the [110] zone axis by varying amounts across the membrane (Figure 4a, top). At higher magnification, the top and bottom membrane boundaries appear visibly curved, accompanied by region-to-region contrast variations indicating local membrane distortion after transfer (Figure 4a, bottom). In contrast, after thermal annealing at 600 ºC, the m-STO becomes significantly flatter showing uniform contrast across a wide field of view (Figure 4b, top), and the corresponding high-magnification image reveals uniform atomic-column contrast, indicating that the membrane is no longer locally tilted (Figure 4b, bottom). These observations show that thermal annealing suppresses bending and improves the structural uniformity of m-STO on $SiO_2$/Si, resulting in a substantially flatter and more conformal membrane-substrate interface. This atomic-scale picture is consistent with the XRD and XRR results (Figure 2d), which reveal reduced mosaicity and interfacial roughness after annealing.

Having established that thermal annealing suppresses structural inhomogeneities in oxide membranes, we next examine whether this structural recovery extends to interfaces in twisted oxide geometries. In such structures, residual interfacial gaps weaken interlayer coupling and may hinder moiré-driven emergent quantum phenomena[30,31,33,39,40]. To investigate this, we transferred an m-STO sample onto an STO substrate with a twist angle of $\phi_{Twist} \approx 11.1^{o}$ and subsequently annealed at 600 °C (Figure S4). Cross-sectional HAADF-STEM shown in Figure 4c reveals a uniform, atomically flat m-STO/STO interface without discernible interfacial gaps. Higher-magnification imaging, together with Sr and Ti elemental maps acquired by energy-dispersive spectroscopy (EDS), resolves the atomic structure and elemental connection across the twisted interface. The corresponding Sr and Ti intensity profiles, extracted along the direction indicated by the white arrow in Figure 4c, show an abrupt transition across the m-STO/STO boundary (Figure 4d), further confirming the formation of a well-defined atomic-scale interface. These results demonstrate that thermal annealing can establish conformal interfacial contact even between crystallographically twisted oxide layers, providing a structural basis for constructing stacked and twisted oxide heterostructures with atomically well-defined interfaces.

**Conclusion and outlook**

In summary, we demonstrate that thermal annealing enables transferred freestanding oxide membranes to recover structural uniformity and form atomically flat, conformal interfaces with the underlying substrate. X-ray diffraction and cross-sectional STEM measurements reveal a substantial reduction in membrane curvature, lattice mosaicity, and interfacial roughness, accompanied by enhanced membrane–substrate contact. This structural recovery restores the intrinsic optical response of STO membranes without introducing detectable low-energy in-gap states. We further demonstrate thermal "ironing" as the critical step to twist-oxide membranes with with no discernible interfacial gaps, thereby, creating atomically conformal interfaces. Such control over membrane–substrate conformity and interfacial uniformity provides an important foundation for stacked, twisted, and heterogeneous oxide architectures, in which well-defined interfaces and structural coherence are critical for enabling emergent quantum phenomena.

## Methods

### Growth of epitaxial oxide heterostructure

Epitaxial $SrTiO_3$ (STO)/$Sr_4Al_2O_7$ heterostructures were grown on (001) STO substrates using pulsed laser deposition. A single-crystalline STO target was used, while for $Sr_4Al_2O_7$ thin films, a sintered polycrystalline $Sr_3Al_2O_6$ target was used. Before deposition, (001)-oriented STO single-crystal substrates (Crystec GmbH, dimensions of $5 \times 5 \times 1$ mm$^3$) were treated with the buffered oxide etcher followed by annealing at 1100 °C to obtain a $TiO_2$-terminated surface. Film growth was performed using a KrF excimer laser (Coherent LPXpro 305 F, $\lambda = 248$ nm) with a laser fluence fixed at 1.0 Jcm$^{-2}$. A 20 nm water-soluble $Sr_4Al_2O_7$ sacrificial layer was first deposited at 750 °C under oxygen partial pressure of 15 mTorr. Subsequently, STO films were grown epitaxially on $Sr_4Al_2O_7$ films at 700 °C and 15 mTorr of oxygen. The growth process was monitored real time using *in situ* reflection high-energy electron diffraction.

### Release and transfer of freestanding $SrTiO_3$ membranes

To provide mechanical support for the membrane during the selective etching of the $Sr_4Al_2O_7$ sacrificial layer in water, 2 μL of a poly(propylene carbonate) (PPC; Sigma-Aldrich) solution, prepared by dissoving 10 g of PPC in 100 mL of anisole, was drop-cast onto the epitaxial heterostructures. The PPC coating was then cured on a hot plate at 70 °C for 10 min. The sample was subsequently immersed in deionized water at room temperature, allowing the $Sr_4Al_2O_7$ sacrificial layer to dissolve selectively. After dissolution, the PPC-supported membranes were detached from the parent substrate and transferred onto a desired substrate. Finally, the PPC support layer was removed by thermal decomposition in air at 250 °C for 1 h, yielding freestanding STO membranes.

### Structural characterization of freestanding oxide membranes

X-ray diffraction (XRD) using a four-circle high-resolution X-ray diffractometer (X'Pert Pro, PANalytical; Cu $K\alpha_1$ radiation) was used to confirm the structural properties of the crystalline films. To investigate the atomic structure of the freestanding membranes, atomic-resolution imaging was perfromed using HAADF-STEM. Cross-sectional STEM lamellae oriented along the STO [110] zone axis were prepared using a focused ion beam (FIB) system (FEI Helios G4). Prior to milling, about 15 nm-thick carbon layer was deposited by electron beam-induced deposition using a gas injection system in the SEM to protect the surface. STEM imaging was carried out using an aberration-corrected, monochromated microscope (Thermo Fisher Scientific Themis Z) operated at an accelerating voltage of 300 kV, with a convergence semi-angle of 18 mrad and a HAADF collection semi-angle of 50-200 mrad. Energy-dispersive X-ray spectroscopy (EDS) was perfromed on the same instrument using a probe current of 100 pA and a pixel dwell time of 2 μs, with acquisition of 300 frames for integration.

**Optical measurement on freestanding $SrTiO_3$ membranes**

**Transmission measurement**

Transmission spectra were obtained using a PerkinElmer Lamda 950, equipped with tungsten-halogen and deuterium lamps and PMT and PbS detectors, covering the spectral range from 175 to 3300 nm. The absorption coefficient was calculated using the membrane thickness and substrate transmission. Samples were mounted on a 1mm pinhole holder.

**Ellipsometry measurement**

Spectroscopic ellipsometry was performed using an M-2000 ellipsometer (J.A. Woollam Co.) to obtain the real ($\varepsilon_1$) and imaginary ($\varepsilon_2$) parts of the dielectric function. Prior to all measurements, the instrument was calibrated at 300 K using a 25 nm $SiO_2$/Si wafer to minimize the mean square error (MSE) of the optical signal. Ellipsometric spectra were acquired at an incident angle of 70 $^{o}$, yielding $\Psi$ and $\Delta$ over the photon-energy range of 0.74 - 6.46 eV.

**Acknowledgments**

This work was sponsored by the U.S. Department of Energy, Office of Science, Basic Energy Sciences, Materials Sciences and Engineering Division. Part of optical spectroscopy and data analysis work at University of Kentucky was supported by the National Science Foundation Grant No. DMR-2426874. Cross-sectional microscopy work was supported by the National Research Foundation of Korea (NRF) grant funded by the Korean government (MSIT) (No. RS-2021-NR060087).

**Author contributions**

J. S. and H. N. L. conceived the idea and designed the experiments; J. S. and H. N. L. performed freestanding membrane fabrication, characterization, and data analysis with the help of T. J. S., Y.-H. K., H. M., and J. Y.; K.-Y. P. and C. S. C. performed the STEM measurements; K. K. and A. S., performed optical transmission measurement and related analysis; J. S. and H. N. L. wrote the manuscript with contributions from all authors.

**Additional information**

**Supplementary Information** is available in the online version of the paper. Correspondence and requests for materials should be addressed to H-. N-. L.

**Data availability**

The data supporting the findings of this study are available from the corresponding author upon request.

**Competing financial interests**

The authors declare no competing financial interests.

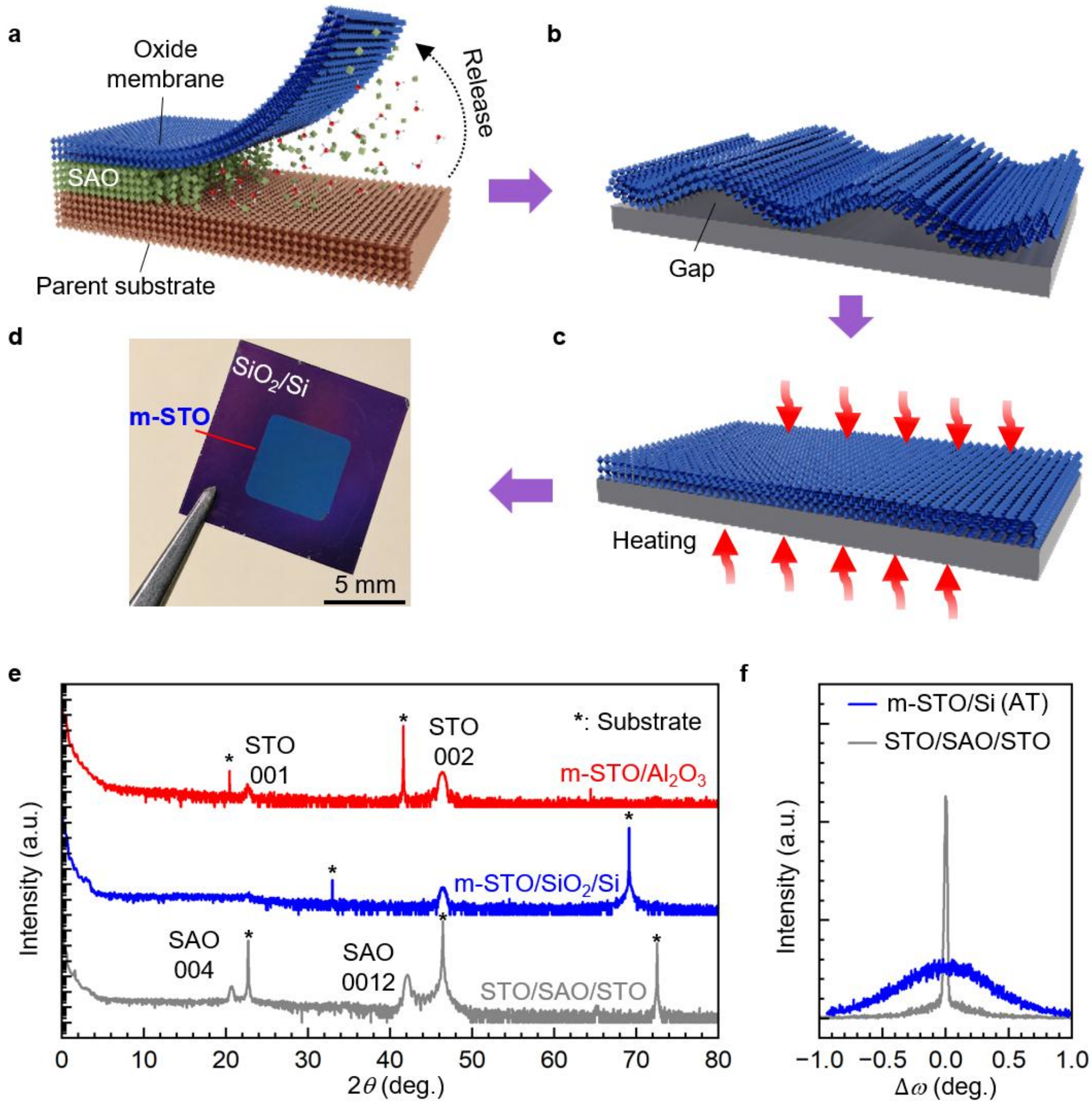

**Figure 1. Fabrication and transfer-induced structural evolution of freestanding STO membranes.** (a-c) Schematic illustration of the sacrifical-layer lift-off process. (a) An epitaxial STO membrane (m-STO) grown on a water-soluble $Sr_4Al_2O_7$ (SAO) sacrificial layer is released by dissolving SAO in water. (b) The membrane is then transferred onto a target substrate. (c) Thermal "ironing" us used to flatten the trasferred membrane. (d) Optical image of an m-STO transferred onto $SiO_2$/Si. (e) X-ray diffraction patterns of the as-grown STO/SAO/STO heterostructure (gray) and m-STOs transferred onto $SiO_2$/Si (blue) and $Al_2O_3$ (red). (f) X-ray rocking curves ($\omega$) of the as-transferred (AT) m-STO (blue) and the as-grown epitaxial STO film (gray), showing a substantial increase in mosaicity after membrane transfer.

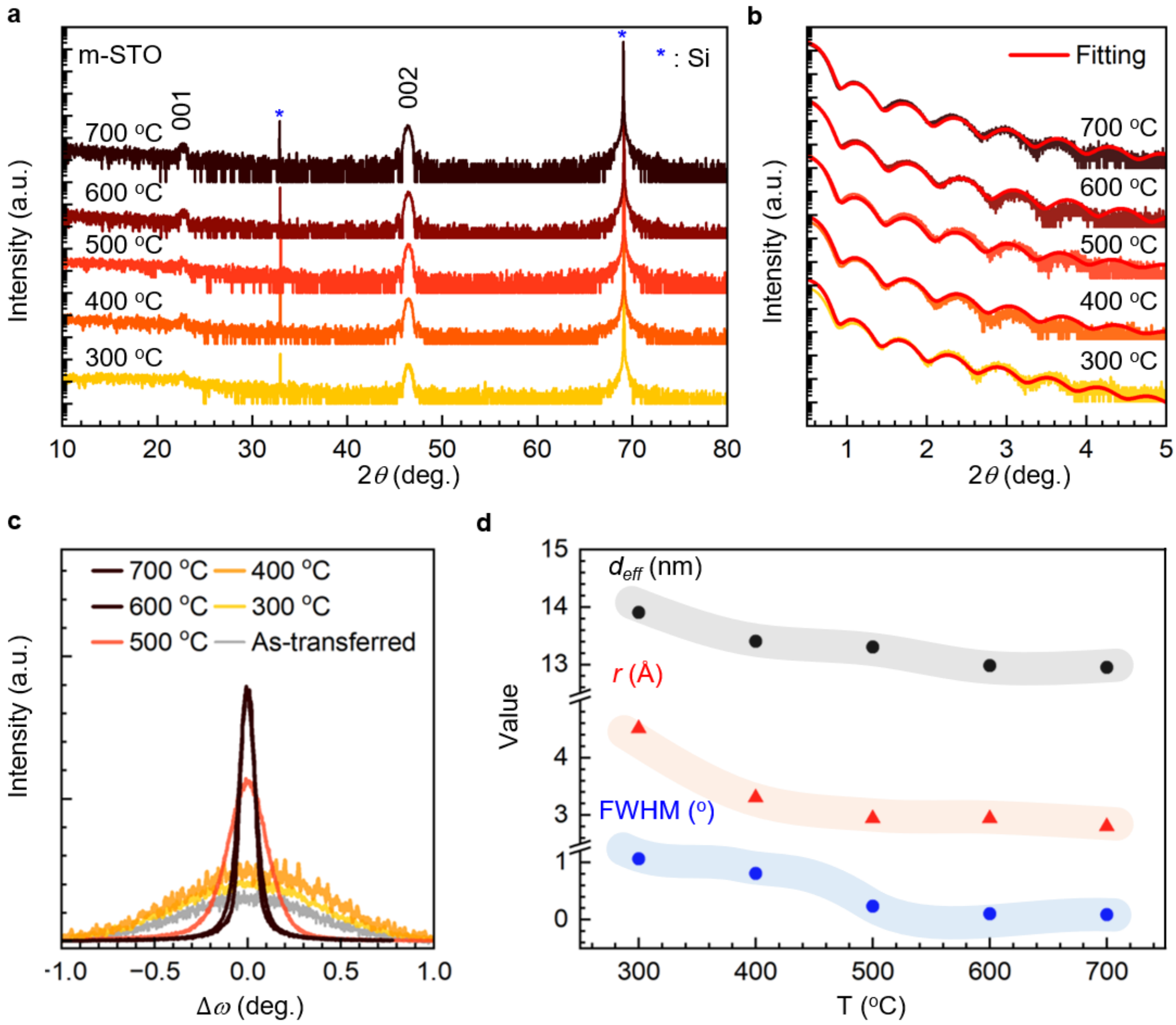

**Figure 2. Structural evolution upon thermal annealing.** (a) XRD 2θ–θ scans of m-STO samples on $SiO_2$/Si after thermal annealing from 300 to 700 °C. The asterisk denotes reflections from the $SiO_2$/Si substrate. (b) Low-angle X-ray reflectivity (XRR) profiles of annealed m-STO, along with corresponding fitting results. (c) XRD rocking curves ($\omega$) of the as-transferred and annealed m-STO, showing progressive narrowing of the peak width with increasing annealing temperature. (d) Extracted membrane thickness ($d_{eff}$), surface/interface roughness ($r$), and full width at half maximum (FWHM) of the rocking curve as a function of annealing temperature. The reduction in roughness and FWHM indicates enhanced membrane flattening and improved structural coherence.

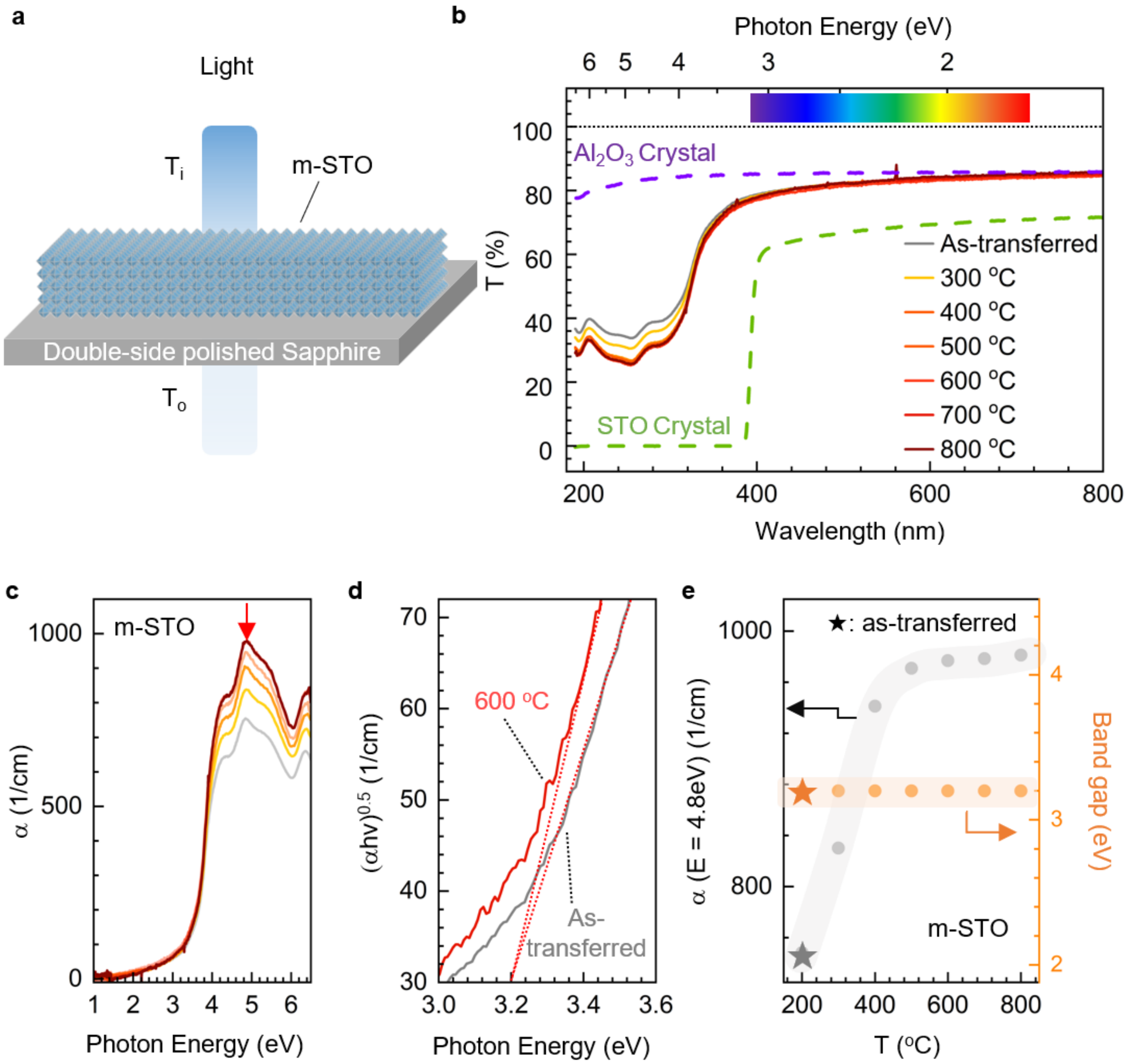

**Figure 3. Optical responses of m-STO following thermal ironing.** (a) Schematic illustration of the optical transmission measurement geometry for an m-STO transferred onto a double-side polished sapphire substrate. Incident light ($T_i$) passes through the m-STO and sapphire substrate, while transmitted light ($T_o$) is collected from the opposite side. (b) Optical transmission spectra of as-transferred and annealed m-STO measured from 200 to 800 nm. The transmission becomes closer to that of single-crystalline STO with increasing the annealing temperature. (c) Extracted absorption coefficient (α) spectra of m-STO. (d) Tauc plots of $(\alpha h\nu)^{0.5}$ versus photon energy for the as-transferred and 600 °C annealed membranes. Linear extrapolation indicates that the optical band gap remains unchanged after annealing. (e) Absorption coefficient at 4.8 eV and extracted optical band gap as a function of the annealing temperature.

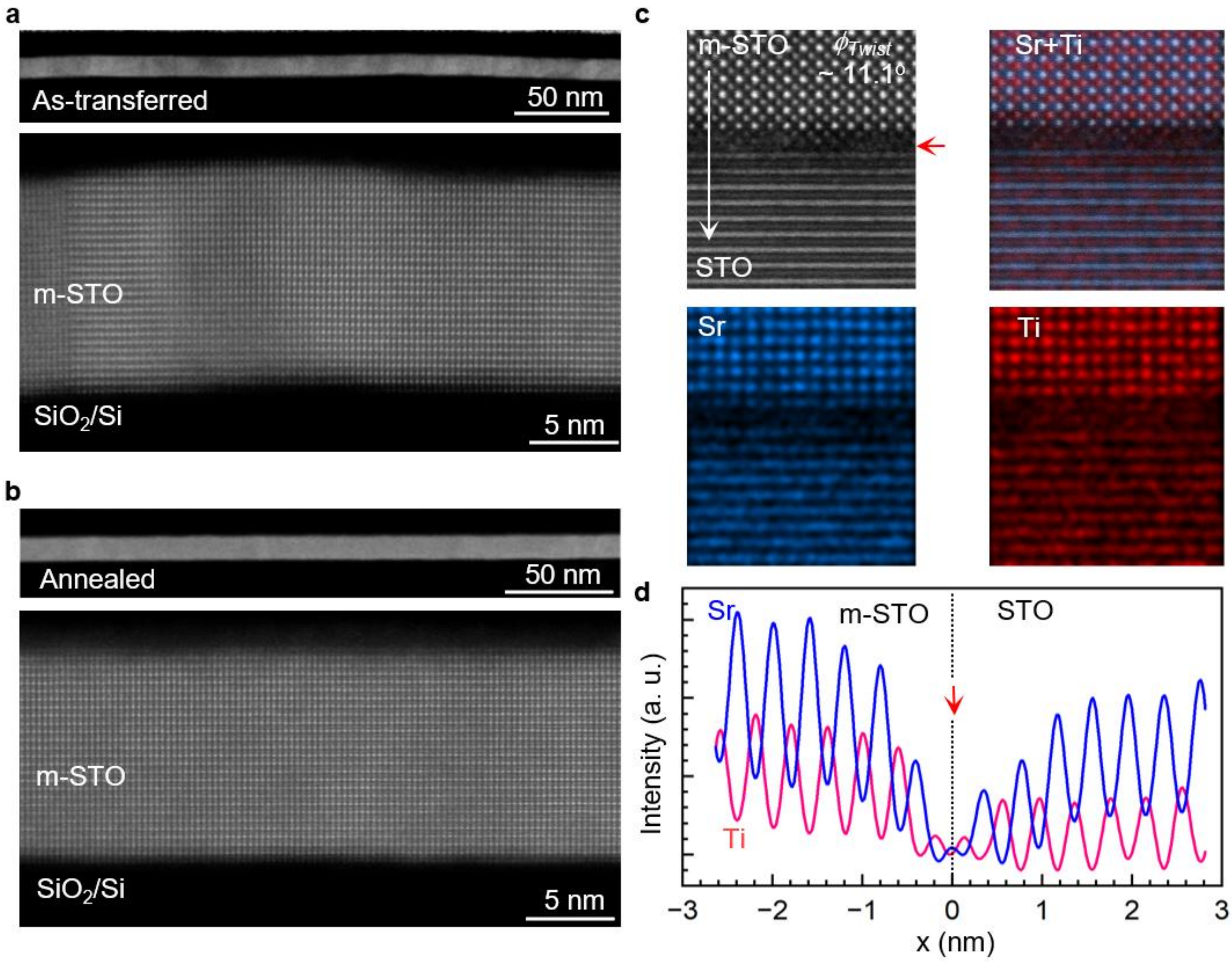

**Figure 4. Cross-sectional STEM analysis of annealing-induced structual enahancements in m-STO on $SiO_2$/Si and at twisted m-STO/STO interfaces.** (a) Low-magnification (top) and high-magnification (bottom) cross-sectional HAADF-STEM images of as-transferred m-STO on $SiO_2$/Si viewed along the STO [110] zone axis, showing region-to-region contrast variations that suggest local membrane bending. (b) Low- magnification (top) and high-magnification (bottom) cross-sectional HAADF-STEM images of annealed m-STO on $SiO_2$/Si. After annealing, the membrane exhibits a flatter morphology with uniform lattice contrast, indicating reduced local bending. (c) HAADF-STEM images and corresponding EDS elemental maps of Sr, Ti, and Sr+Ti for an annealed, twisted m-STO sample on an STO substrate with a twist angle ($\phi_{Twist}$) of ~11.1°. The EDS maps show a uniform, atomically flat interface without gaps. (d) Ti and Sr intensity profiles acquired across the interface along the region indicated in (c). The red arrow marks the m-STO/STO interface**.**